\documentclass{article}

\usepackage{arxiv}

\usepackage[utf8]{inputenc} 
\usepackage[T1]{fontenc}    
\usepackage{hyperref}       
\usepackage{url}            
\usepackage{booktabs}       
\usepackage{amsfonts}       
\usepackage{nicefrac}       
\usepackage{microtype}      
\usepackage{lipsum}		
\usepackage{graphicx}
\usepackage{natbib}
\usepackage{doi}
\usepackage{booktabs}
\usepackage{multirow}
\usepackage{xcolor,colortbl}

\DeclareMathAlphabet\mathbb{U}{msb}{m}{n}
\usepackage{enumitem}
\usepackage[vlined,ruled,linesnumbered]{algorithm2e}
\usepackage{nth}
\graphicspath{{Fig/}}
\usepackage{hyperref}
\usepackage{xcolor}

\usepackage[most]{tcolorbox}
\newtcolorbox{highlighted}{colback=yellow,coltext=black,breakable}
\usepackage{fdsymbol}
\usepackage[leqno]{mathtools}
\usepackage{graphicx}
\usepackage{color}
\usepackage{amsmath,amsfonts}
\usepackage{mathrsfs}
\usepackage{hyperref}
\usepackage{wrapfig}
\usepackage{listings}
\usepackage{ifpdf}
\hypersetup{pdfauthor={Name}}
\usepackage{subfig}
\usepackage{booktabs}
\newtheorem{definition}{Definition}
\newtheorem{example}{Example}

\newcommand{\f}{\varphi}

\newcommand{\Intvl}{I}
\newcommand{\sPara}{c}

\newcommand{\val}{\nu}

\newcommand{\x}{\mathbf{x}}
\newcommand{\F}{\mathbf{F}}
\newcommand{\true}{\mathit{true}}
\newcommand{\aand}{\,\wedge\,}

\newcommand{\Reals}{\mathbb{R}}

\newcommand{\PosReals}{\mathbb{R}^{\ge 0}}
\newcommand{\domain}{\mathcal{D}}
\newcommand{\setof}[1]{\left\{#1\right\}}

\newcommand{\mypara}[1]{\vspace{0.3em} \noindent {\bf #1.\ }}

\newcommand{\timedomain}{\mathbb{T}}

\newcommand{\framework}{\textsc{DialogueSTL}}
\newcommand{\demo}{\mathbf{d}}
\newcommand{\state}{\mathbf{s}}
\newcommand{\action}{\mathbf{a}}
\newcommand{\alw}{\mathbf{G}}
\newcommand{\Until}{\mathbf{U}}
\newcommand{\atom}[1]{\textit{#1}}
\newcommand{\oor}{\,\vee\,}

\title{\LARGE \bf
Interactive Learning from Natural Language\\and Demonstrations using Signal Temporal Logic}

\author{ Sara Mohammadinejad \\
	University of Southern California\\
	LA, CA, USA \\
	\texttt{saramoha@usc.edu} \\
	\And
	Jesse Thomason\\
	University of Southern California\\
	LA, CA, USA \\
	\texttt{jessetho@usc.edu} \\
	\And
	Jyotirmoy V. Deshmukh\\
	University of Southern California\\
	LA, CA, USA \\
	\texttt{jdeshmuk@usc.edu} \\
}

\renewcommand{\shorttitle}{Learning Logical Formulas from Natural Language}

\hypersetup{
pdftitle={Interactive Learning from Natural Language and Demonstrations using Signal Temporal Logic},
pdfsubject={q-bio.NC, q-bio.QM},
pdfauthor={Sara Mohammadinejad, Jesse Thomason, Jyotirmoy V. Deshmukh},
pdfkeywords={Natural Language Processing, Human Robot Interactions, Signal Temporal Logic},
}

\begin{document}
\maketitle

\begin{abstract}
Natural language is an intuitive way for humans to communicate tasks to a robot.
While \textit{natural language} (NL) is ambiguous, real world tasks and their safety requirements need to be communicated unambiguously.
Signal Temporal Logic (STL) is a formal logic that can serve as a versatile, expressive, and unambiguous \textit{formal} language to describe robotic tasks. 
On one hand, existing work in using STL for the robotics domain typically requires end-users to express task specifications in STL -- a challenge for non-expert users.
 On the other, translating from NL to STL specifications is currently restricted to specific fragments.
In this work, we propose \framework{}, an interactive approach for learning correct and concise STL formulas from (often) ambiguous NL descriptions. 
We use a combination of semantic parsing, pre-trained transformer-based language models, and user-in-the-loop clarifications aided by a small number of user demonstrations 
to predict the best STL formula to encode NL task descriptions. An advantage of mapping NL to STL is that there has been considerable recent work on the use of reinforcement learning (RL) to identify control policies for robots. 
We show we can use Deep Q-Learning techniques to learn optimal policies from the learned STL specifications. 
We demonstrate that \framework{} is efficient, scalable, and robust, and has high accuracy in predicting the correct STL formula with a few number of demonstrations and a few interactions with an oracle user.

\end{abstract}



\section{Introduction}
Natural language is an easy and end-user friendly way for humans to convey their intended tasks to robots. 
However, natural language descriptions are usually ambiguous and can have multiple meanings. 
For example, “pick up the door key and open the door” implicitly carries urgency to a human if there is a fire in the room.
However, a robot system needs to infer timing constraints, as well as underspecified information like \textit{which} door when there are multiple and whether pausing between getting and using the key is acceptable.

Signal Temporal Logic (STL) \cite{maler2004monitoring} has been used as a flexible, expressive, and unambiguous language to describe robotic tasks that involve time-series data and signals.
For instance, STL can be used to formulate properties such as “the robot should immediately extinguish fires but can accept delays in opening doors.”
From a grammar-based perspective, an STL formula can be viewed as atomic formulas combined with logical and temporal operators \cite{mohammadinejad2020interpretable}.
In this work, we match different components of a natural language description with atoms and operators to form candidate STL formulas, and use dialogue with users to resolve ambiguities.

Formalizing behavior using temporal logics such as STL require the user to specify the correct temporal logic specification~\cite{balakrishnanstructured,puranic2021learning}---a difficult and error-prone task for untrained human users.
Translating a sentence written in a natural and ambiguous language into a more general and concise formal language is an open challenge~\cite{he2021english}.

\begin{figure}[t]
\centering
\includegraphics[scale=0.45]{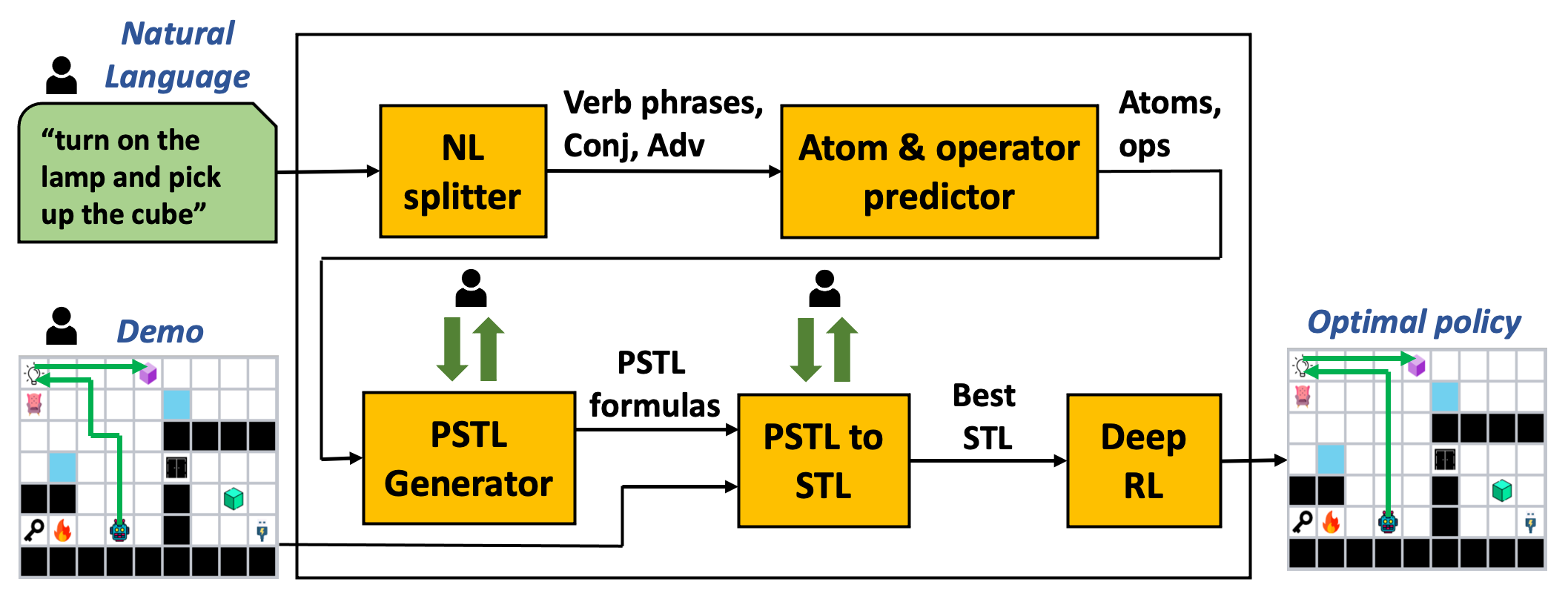}
\caption{We infer a STL formula and optimal policy from a given natural language description, a few demonstrations, and questions to the user.}
\label{fig:highlevel}
\end{figure}

Seminal works in \cite{gavran2020interactive} considers the problem of interactive synthesis of robot policies from natural language specifications. The authors consider similar environments as ours but expect specifications to be provided in structured English that is then translated to LTL formulas. The main objective is to explore the space of specifications using user demonstrations and constraint-based methods to identify the precise specifications, which are then used to synthesize robot policies using reactive synthesis methods. In our work, the emphasis is on directly learning the structure of the task objective using modern natural language processing tools and employ a dialogue-based method to refine the specification. Furthermore, we use recently developed RL methods to learn optimal policies from STL instead of LTL synthesis approaches.


Prior work closest to ours, DeepSTL~\cite{he2021english}, translates informal English requirements to STL by training a sequence-to-sequence transformer on \textit{synthetic} data generated from a hand-defined STL-to-English grammar.
DeepSTL is restricted to a specific fragment of STL covered by the hand-defined grammar.
For example, it allows the conjunction and disjunction of only two atomic propositions, and some nested formulas are not supported.
Further, DeepSTL directly translates natural language into STL without any interaction with the user, meaning language that is missing required information for the target STL simply cannot be correctly understood.
Natural language is vague, and we propose clarifying interactions to arrive at the correct STL formula.

We propose an interactive and flexible approach, \framework{}.
Figure.~\ref{fig:highlevel} shows the high-level flow of our framework, whose input consists of natural language description (NL) and a few demonstrations (demos) of a robotic task. 
After clarifying ambiguities with the user via dialogue, our framework produces the optimal policy as outcome. 
We demonstrate that our method is efficient and scalable, and note that in most cases the user has to provide only a few demonstrations---often only one---of a successful behavior for our framework to arrive at the correct STL formula using dialogue interactions.



\section{Background}
\label{sec:background}
\begin{definition}[Demonstration]
Demonstration is a finite sequence of state-action pairs. Formally, $\demo=\{(\state_0, \action_0), (\state_1, \action_1),...,(\state_\ell, \action_\ell)\}$ defines a demonstration with length $\ell$, where $\state_i \in S$ and $\action_i \in A$. $S$ is the set of all possible states and $A$ is the set of all possible actions in an environment.
\end{definition}

\begin{definition}[Time-Series, Traces, Signals]
A trace $\x$ is a mapping from time domain $\timedomain$ to value
domain $\domain$, $\x: \timedomain \rightarrow \domain$ where,
$\timedomain \subseteq \PosReals$, $\domain \subseteq \Reals^n$, $ \timedomain \neq \emptyset$, and the variable n denotes the trace dimension.
\end{definition}

\mypara{Signal Temporal Logic (STL)}
Signal Temporal Logic (STL) is a logic to reason about properties of real-valued signals. 
The basic primitive in STL is called atomic predicate or atom. 
Atoms are formulated as $f(\x) \sim \sPara$,
where $\x$ is a trace, $f$ is a scalar-valued function
over the trace $\x$, $\sim \;\in
\setof{\geq, \leq,=}$, and $\sPara \in \Reals$. For instance, $\x
\leq 1$ is an atomic predicate, where $f(\x) = \x$, $\sim$ is $\leq$,
and $c=1$. 
Temporal specifications are created by adding operators such as $\alw$
(always), $\F$ (eventually) and $\Until$ (until) to atoms.
For example, $\alw (\x \leq
1)$ means that signal $\x$ is \textit{always} less than or equal to 1. Each temporal
operator is indexed by an interval $\Intvl := (a,b) \mid (a,b] \mid
[a,b) \mid [a,b]$, where $a, b \in \timedomain$, and $\timedomain \in \Reals$ is the time domain. 

Formally,
\begin{equation}
\label{eq:stl_syntax}
\begin{array}{l}
\f :=      \true 
      \mid f(\x) \sim \sPara 
      \mid \neg\f \mid \f_{1} \wedge \f_{2} 
      \mid \f_{1}\, \Until_{\Intvl}\, \f_{2},
\end{array}
\end{equation}
where $c \in \Reals$. $\alw$ and $\F$ operators are special instances of
$\Until$ operator \mbox{$\F_{\Intvl}\varphi \triangleq \true\, \Until_\Intvl\,
\varphi$}, and \mbox{$\alw_{\Intvl}\varphi \triangleq \neg \F_{\Intvl}
\neg \varphi$}.

The Boolean satisfaction of an atomic predicate is \atom{true} if the predicate is satisfied and \atom{false} if it is violated, and the semantics of logical and temporal operators are defined as:

\begin{itemize}
	\item $\neg \varphi$ is satisfied if $\varphi$ is not satisfied or $\varphi$ is violated.
	\item $\varphi_1 \land \varphi_2$ holds if both $\varphi_1$ and $\varphi_2$ are satisfied.
	\item $\varphi_1 \lor \varphi_2$ holds if either $\varphi_1$ or $\varphi_2$ is satisfied.
	\item $\varphi_1 \rightarrow \varphi_2$ is equivalent to $\neg \varphi_1  \lor \varphi_2$, which means that either  $\varphi_1$ should not hold or $\varphi_2$ should hold.
    \item $\alw_I\varphi$ means $\varphi$ must hold for all instances of interval $I$.
    \item $\F_I\varphi$ means $\varphi$ must hold at
    least once in interval $I$.
    \item $\varphi_1 \Until_I \varphi_2$ means $\varphi_1$ must hold in $I$ until
    $\varphi_2$ is satisfied.
\end{itemize}

\begin{figure}[t]
\centering
\includegraphics[scale=0.45]{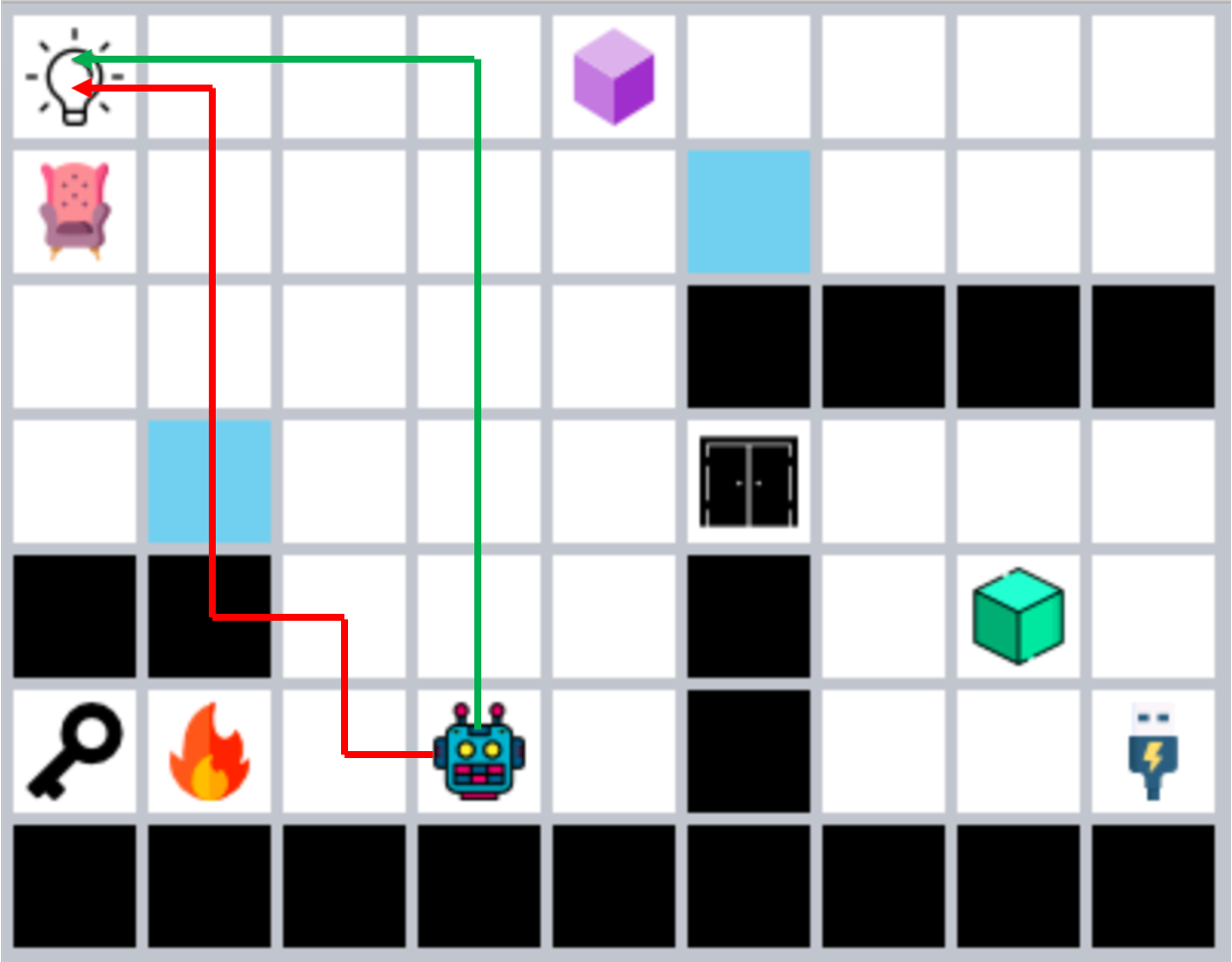}
\caption{The robot tries to reach the lamp placed at location $(0,0)$ in 15 seconds while avoiding wall (black) and water (blue) tiles.
Both green (\textcolor{green}{---}) and red (\textcolor{red}{---}) demonstrations satisfy the formula $\F_{[0,15]}(\atom{robotAt(0,0)})$; in the next 15 seconds, the robot should eventually reach to the location $(0,0)$.
}
\label{fig:grid}
\end{figure}

\begin{example}
Consider the grid world environment illustrated in Fig.~\ref{fig:grid}. 
While both demonstrations reach the lamp, only the green demonstration satisfies the formulas $\alw(\neg(\atom{robotAtWall}))$ and $\alw(\neg(\atom{robotAtWater}))$. 
The formula $\alw(\neg(\atom{robotAtWall}))$ means that the robot should never climb walls. 
The formula $\alw(\neg(\atom{robotAtWater}))$ means that the robot should not step in water.
The red demonstration intersects with both walls and water tiles.
\end{example}

\mypara{Parametric Signal Temporal Logic (PSTL)}A PSTL \cite{asarin2011parametric} formula
is an extension of STL formula where constants are replaced by
parameters. A STL formula is obtained by pairing a PSTL formula with a
valuation function that assigns a value to each parameter variable.
For example, consider the PSTL formula $\f(x,y, \tau) = \F_{[0,\tau]}(\atom{robotAt(x,y)})$ with parameters $\tau$, $x$ and $y$. The STL formula $\F_{[0,15]}(\atom{robotAt(0,0)})$, which is an instance of $\f$, is obtained with the valuation
$\val = \setof{\tau \mapsto 15, x \mapsto 0, y \mapsto 0}$. From a grammar-based perspective, a PSTL formula can be viewed as
atomic formulas combined with unary or binary operators \cite{mohammadinejad2020interpretable}.
\begin{equation}
\begin{array}{l}
\f := atom \mid unaryOp(\f) \mid binaryOp(\f, \f) \\
unaryOp := \neg \mid \F \mid \alw\\
binaryOp := \vee \mid \wedge \mid \Until_I \mid \Rightarrow 
\end{array}
\end{equation}

For instance, PSTL formula $\F_{[0,\tau]}(\neg \atom{robotAt(a,b)} \oor \atom{robotAt(c,d)})$ consist of atoms $\atom{robotAt(a,b)}$ and $\atom{robotAt(c,d)}$, unary operators $\neg, \F$ and binary operator $\oor$.

\section{\framework{}: Learning PSTL Candidates}
\label{sec:nl_to_pstl}
\newcommand{\taskNL}{\mathtt{taskNL}}
\newcommand{\trainn}{\mathtt{train}}
\newcommand{\testt}{\mathtt{test}}
\newcommand{\atomm}{\mathtt{atom}}
\newcommand{\atoms}{\mathtt{atoms}}
\newcommand{\sampleAtoms}{\mathtt{sampleAtoms}}
\newcommand{\trainAtoms}{\mathtt{trainAtoms}}
\newcommand{\testAtoms}{\mathtt{testAtoms}}
\newcommand{\assigns}{\coloneqq}
\newcommand{\vPhrases}{\mathtt{vPhrases}}
\newcommand{\vPhrase}{\mathtt{vPhrase}}
\newcommand{\conjs}{\mathtt{Conjs}}
\newcommand{\advs}{\mathtt{Advbs}}
\newcommand{\taggedTokens}{\mathtt{taggedTokens}}
\newcommand{\sampleDataOps}{\mathtt{sampleOps}}
\newcommand{\op}{\mathtt{op}}
\newcommand{\ops}{\mathtt{ops}}
\newcommand{\confidence}{\mathtt{confidence}}
\newcommand{\opEmbeddings}{\mathtt{opEmbeddings}}
\newcommand{\PSTLFormulas}{\mathtt{PSTLFormulas}}
\newcommand{\lowerr}{\mathtt{l}}
\newcommand{\upper}{\mathtt{u}}

\begin{figure*}[t]
\centering
\includegraphics[scale=0.65]{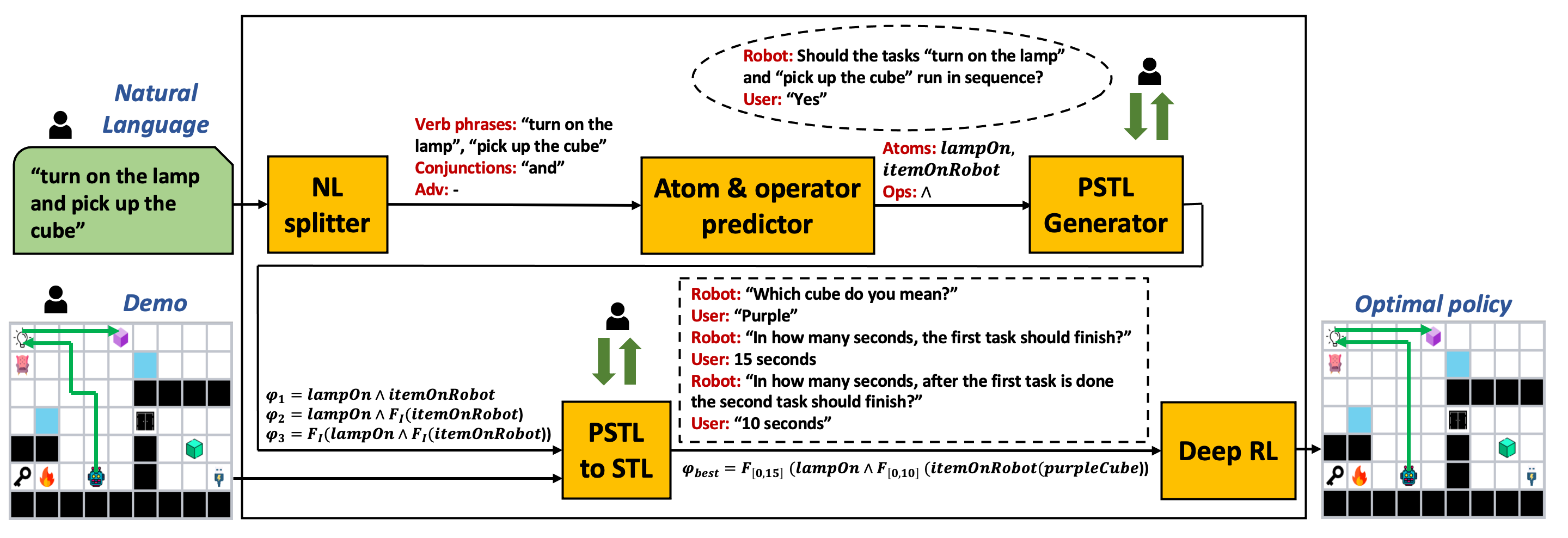}
\caption{We infer a STL formula $\varphi_{best} = \F_{[0,15]}(\atom{lampOn} \wedge \F_{[0,10]}(\atom{itemOnRobot(purpleCube)}))$ and an optimal policy from a given natural language description ``Turn on the lamp and pick up the cube", a demonstration (\textcolor{green}{---}), and interactions with the user. 
NL splitter extracts components of the Natural Language (i.e., ``Turn on the lamp", ``and", ``pick up the cube"). 
Each component is mapped to an atom or operator using the Atom and Operator Predictors (``Turn on the lamp":$\atom{lampOn}$, ``and":$\wedge$, ``pick up the cube":$itemOnRobot$). 
Next, candidate PSTL formulas are generated from the predicted atoms and operators. 
Asking questions from the user can help learn parameters of PSTL formulas and therefore, learning the best STL formula. 
Finally, Deep RL techniques are employed to learn an optimal policy from the learned STL formula.}
\label{fig:running_example}
\end{figure*}

In this section, we propose an interactive approach to learn candidate Parametric STL (PSTL) formulas from the natural language description of a task or constraints provided by user. The overall structure of our approach is shown in Algo.~\ref{alg:nl_to_pstl}. 
As a running example, consider the command ``turn on the lamp and pick up the cube'' for the grid world environment shown in Fig.~\ref{fig:grid}.
The high-level view of our method for the running example is illustrated in Fig.~\ref{fig:running_example}, and the method for converting NL to candidate PSTL formulas is formalized in Algo.~\ref{alg:nl_to_pstl}.
The inputs of the algorithm consist of NL description of the task ($\taskNL$), a few sample data for each atomic predicate ($\sampleAtoms$) and operator ($\sampleDataOps$) that we generate manually, and a threshold $\epsilon$ on the confidence of an atom predictor. 
We first generate a set of data for atom predictor (AtomPredictor) using a GPT-3 based paraphrase generator (GPT3ParaphGen) and learn a model for predicting likely atoms from individual verb phrases.\footnote{We denote a \textit{verb phrase} loosely as a group of words that contains a verb, such as ``turn on the lamp," ``if fire is on," and ``open the door."} 
For a given NL description, verb phrases are extracted and matched with atoms and operators, and candidate PSTL formulas with length in the range $[l,u]$ are enumerated using predicted atoms and operators. We now explain each part of Algo.~\ref{alg:learning_best_stl} step-by-step.

\DontPrintSemicolon

\begin{algorithm}
\small
\caption{Natural Language to PSTL algorithm \label{alg:nl_to_pstl}}
\SetKwProg{Fn}{Function}{:}{}
\KwIn{\textit{$\taskNL$, $\sampleAtoms$, $\sampleDataOps$, $\epsilon$=0.3}}
\KwOut{$\PSTLFormulas$}
{\color{teal}\tcp{Generate data for atom predictor}}
$\atoms$ $\assigns$ GPT3ParaphGen($\mathtt{sampleAtoms}$) \;
$\trainAtoms$, $\testAtoms$ $\assigns$ TrainTestSplit($\atoms$) \;
AtomPredictor $\assigns$ $\trainn(\trainAtoms)$ \tcp*[r]{Train}
Accuracy $\assigns$ $\testt(\testAtoms)$ \tcp*[r]{Test} 
{\color{teal}\tcp{Avg. embedding for each operator}}
$\opEmbeddings$ $\assigns$ computeOpsBertEmbeddings($\sampleDataOps$) \;
$\taggedTokens$ $\assigns$ partOfSpeechTagger($\taskNL$) \;
{\color{teal}\tcp{Extract verb phrases, conjunctions and adverbs based on the tags}}
$\vPhrases, \conjs, \advs$ $\assigns$ NLSplitter($\taggedTokens$) \;
\For(\tcp*[f]{Find best atoms}){$\vPhrase \in \vPhrases$}{
$\atomm, \confidence$ $\assigns$ AtomPredictor($\vPhrase$)
\If{$\confidence \leq \epsilon$}{
 $\vPhrase$ $\assigns$ ParaphrasedByUser($\vPhrase$) \;
 $\atomm, \confidence \assigns$ AtomPredictor($\vPhrase$) 
}
}

{\color{teal}\tcp{Find best operators}}
$\ops$ $\assigns$ findBestOps($\opEmbeddings$, $\conjs$, $\advs$)\;
{\color{teal}\tcp{Bounds on the length of PSTL formula}}
$\lowerr,\upper \assigns 2\! \cdot\! |\vPhrases|\! -\! 1, 2\!\cdot\! |\vPhrases|\! +\! |\conjs|\! +\! |\advs|$\;
{\color{teal}\tcp{Enumerative, interactive PSTL synthesis}}
$\PSTLFormulas \assigns$ genPSTLEnum$(\atoms, \ops, \lowerr, \upper)$\;

\end{algorithm}

\mypara{Atom Predictor Data} We overcome the brittleness of a language-to-STL predictor based on hand-crafted grammar~\cite{he2021english} using a state-of-the-art paraphrase generation tool for data augmentation over a small manual set.

GPT-3 \cite{brown2020language} is a large, pretrained language model
able to generate natural language texts that can be hard to distinguish from human-authored content~\cite{dale2021gpt}.
We prime GPT-3 with a few examples such as ``hi" and ``hello" to establish the paraphrase task.
Then, we input manually generated verb phrases corresponding to STL atoms in our grid world. 
We check the paraphrase quality of the GPT-3 outputs manually. 
For the grid world demonstrated in Fig.~\ref{fig:grid}, there are a total of 15 atoms, and we generate a total of 81 verb phrase samples for the 15 atoms.
For example, our GPT-3-based paraphrase generator gets ``Turn off the fire" as input and generates ``Extinguish the fire" as one output paraphrase, and both can be paired with the atom $\atom{fireOff}$ as training data.

\mypara{Atom Predictor} 
Given a set of verb phrases and their corresponding atomic formulas from the aforementioned data generation, we learn a model that given a verb phrase can output the most similar atom to the verb phrase. 
We reserve $80\%$ of the data generated by GPT-3 for training and $20\%$ for validation. 
We use DIET~\cite{bunk2020diet}, a lightweight transformer-based architecture that outperforms fine-tuning BERT~\cite{devlin2018bert} and is about six times faster to train. 
We trained DIET for 100 epochs, which took 14 seconds and resulted in training accuracy of $100\%$ and validation accuracy of $93\%$.

\mypara{Operator Predictor} We compute an average word embedding for each operator. 
To do so, we map each operator with a few words in natural language that correspond to that operator. 
For example, the words ``and'' and ``and then" correspond to $\wedge$ operator.
The word embedding of each operator is computed as the average BERT embedding of the words corresponding to that operator. The computed word embedding will be used to match each conjunction or adverb with the most similar operator.

\mypara{Natural Language Splitting}
To extract verb phrases, we first run a part-of-speech tagger, Flair. 
Flair~\cite{akbik2019flair} is a state-of-the-art tagger based on contextual string embeddings and neural character language models.
We divide the language description based on the position of verbs, resulting in, for example, ``turn on the lamp" and ``pick up the cube". 
We extract conjunctions from the words that connect the verb phrases, such as ``and''.
We also extract any adverbs, since, for example, ``always" can correspond to the globally operator $\alw$.
We try to match each verb phrase with an atom using the trained atom predictor and each conjunction or adverb with an operator using cosine similarity between the operators' and words' BERT embeddings.
We always add $\F$ to the list of candidate operators because it is a common operator. For our running example, we extract $atoms = [\atom{lampOn}, \atom{itemOnRobot}]$ and $operators=[\aand, \F]$. 
If the verb phrase to atom correspondence confidence is low, we ask user to paraphrase the word sequence that has low confidence. 

\mypara{PSTL Generation} We must bound the lower and upper lengths of possible PSTL formulas to enumerate possibilities. 
We consider the lower bound as $l = |\atom{vPhrases}| + |\atom{vPhrases}| - 1$ because $n$ verb phrases need $n-1$ connectors. 
We consider  $u = 2 \cdot |\atom{vPhrases}| + |\atom{Conj}| + |\atom{Adv}|$ as the length upper bound.
We multiply $|\atom{vPhrases}|$ by 2 because each verb phrase can require a $\mathbf{F}$ operator. 
Each conjunction or adverb might also be converted to an operator. 
Next, we use systematic PSTL enumeration \cite{mohammadinejad2020interpretable} to generate candidate PSTL formulas within the range $l$ and $u$ using the extracted atoms and operators in increasing order of their length.
We remove enumerated PSTL formulas that do not contain all the atoms or contain more than one instance from each atom. 

\mypara{Causal and Temporal Dependency}
We shrink the space of candidate PSTL formulas using the idea of causal or temporal dependency between atoms.
$Atom_2$ is causally dependent on $Atom_1$ if $Atom_1$ should happen before $Atom_2$, which eliminates formulas such as $\F(Atom_1) \aand \F(Atom_2)$ and $Atom_2 \aand \F(Atom_1)$.

For our running example, we ask the user whether the tasks ``turn on the lamp" and ``pick up the cube" should run in sequence or not. 
Alternatively, we could ask user if ``turn on the lamp and then pick up the cube" is acceptable or not.
If the answer is ``Yes'', this means that the atom $\atom{itemOnRobot}$ is causally dependent on $\atom{lampOn}$, and hence, formulas $\F(\atom{lampOn}) \aand \F(\atom{itemOnRobot})$ and $\atom{itemOnRobot} \aand \F(\atom{lampOn})$ would be omitted. 
For our running example, the reason for causal dependency is that if room is dark, it would be difficult for the robot to identify the cube and pick it up. 
From the 9 generated PSTL formulas 6 of them are removed resulting in 3 remaining candidate PSTL formulas.  
\begin{align*}
&\varphi_1 = \atom{lampOn} \wedge \atom{itemOnRobot}\\
&\varphi_2 = \atom{lampOn} \wedge \F_\Intvl(\atom{itemOnRobot})\\
&\varphi_3 = \F_\Intvl(\atom{lampOn} \aand \F_\Intvl(\atom{itemOnRobot}))\\
\end{align*}
\section{\framework{}: Selecting Correct STL}
\label{sec:learning_best_stl}
We find the parameters of the PSTL candidates by searching in the initial language description and by interacting with the user (Algo.~\ref{alg:learning_best_stl}).
We use 4 question types: order of tasks, atom parameters, operator parameters, and for paraphrasing a verb phrase.

\framework{} takes positive and negative demonstrations as input.
Asking the user to generate many demonstrations makes for a tedious interface.
We use only a few demonstrations from the user and use them to automatically generate more positive and negative demonstrations.

To generate negative examples, we use the principle of ``no excessive effort''~\cite{gavran2020interactive}. 
Any prefix of the demonstrated positive example is assumed insufficient and considered a negative example.
Intuitively, if a prefix of $\demo$ would already be a good example, then the user would not have given the full demo $\demo$. 
We also add delays between different actions and asking the user to categorize whether each is a positive or negative example. 

To convert each PSTL formula to its corresponding STL formula, all parameter values must be discovered. 
For instance, operator $\F$ needs a time interval and atom $itemOnRobot$ requires the name of the item that should be picked up. 
We search for parameter values in NL description, and if any of the parameter values cannot be found, we find it by interaction with user. 
We replace the parameter valuations in PSTL formulas which result in STL candidates. 
Finally, we choose the STL formula that satisfies positive demos and does not satisfy the negative demos. 
In our running example, among the three candidate STL formulas, $\varphi_3$ is chosen as the best STL formula.

If no STL formula can be found, it may be that the language description was not correctly split into its components, or that one or more of the predicted atoms and operators were not correct. 
In the future, we can improve the splitting algorithm by learning from incorrect predictions, and use beam search in enumeration to move to the next high rank atom or operator when the first set fails.

\newcommand{\demos}{\mathtt{Demos}}
\newcommand{\atomParams}{\mathtt{atomParams}}
\newcommand{\opParams}{\mathtt{opParams}}
\newcommand{\found}{\mathtt{found}}

\begin{algorithm}
\small
\caption{Learning the best STL formula\label{alg:learning_best_stl}}
\SetKwProg{Fn}{Function}{:}{}
\KwIn{\textit{$\taskNL$, $\atoms$, $\ops$, $\PSTLFormulas$, $\demos$ $\demo$}}
\KwOut{\textit{$\varphi_{best}$}}
{\color{teal}\tcp{Generate more positive, negative demos}}
$\demo^+, \demo^-$ $\assigns$ generateMoreDemos($\demo$) \;
{\color{teal}\tcp{Finding atoms' parameters}}
\For{$\atomm \in \atoms$}{
$\atomParams$ $\assigns$ findAtomParams($\taskNL$, $\atomm$)\\
\If{$\atomParams\;\textbf{not}\;\found$}{
$\atomParams$ $\assigns$ getParamsByInteractionWithUser()
}
}
{\color{teal}\tcp{Finding operators' parameters}}
\For{$\op \in \ops$}{
$\opParams$ $\assigns$ findOpParams($\taskNL$, $\op$)\\
\If{$\opParams\;\textbf{not}\;\found$}{
$\opParams$ $\assigns$ getParamsByInteractionWithUser()
}
}
\For{$\varphi(p)$ $\in$ $\PSTLFormulas$}{
{\color{teal}\tcp{Set parameters of the PSTL formula}}
$\varphi(v(p))$ $\assigns$ setParams($\varphi(p)$, $\atomParams$, $\opParams$)\\
{\color{teal}\tcp{Check if the STL formula $\varphi(v(p))$ satisfies positive demos and does not satisfy negative demos}}
\If{$\varphi(v(p)) \vDash  \demo^+$ and $\varphi(v(p)) \nvDash  \demo^-$}{
$\varphi_{best} \assigns \varphi(v(p))$\\
$\Return\;\varphi_{best}$
}
}
$\Return\;\emptyset$
\end{algorithm}

\section{Learning optimal policies}
\label{sec:rl}
\begin{figure}[t]
\centering
\includegraphics[scale=0.5]{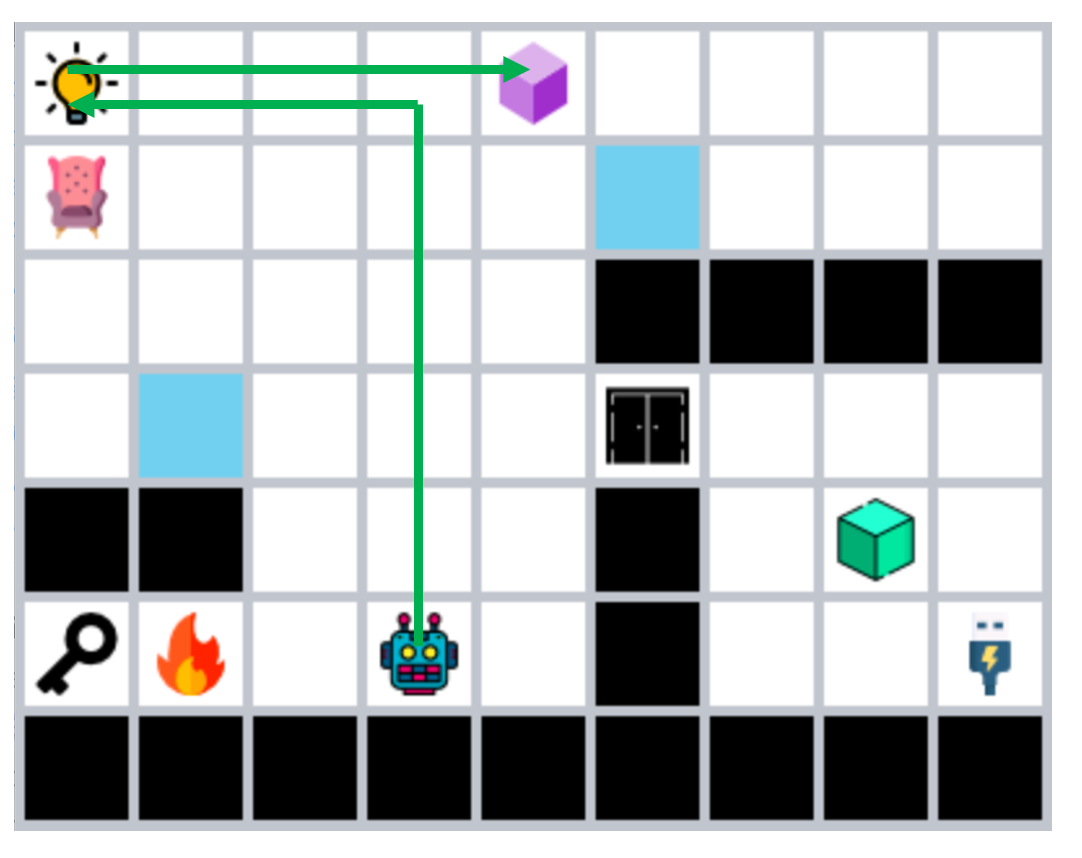}
\caption{The learned policy (\textcolor{green}{---}) from the $\varphi_{task} = \F_{[0,15]}(\atom{lampOn} \aand \F_{[0,10]}(\atom{itemOnRobot(purpleCube)}))$.}
\label{fig:policy}

\end{figure}

Previous works have used the robustness of an STL formula, or the signed distance of a given trajectory from satisfying or violating a given formula as rewards to guide the RL algorithm~\cite{puranic2021learning, balakrishnanstructured}.
Here, we only provide an example of learning optimal policy from a given STL formula using those existing techniques.
We use Deep Q-Learning~\cite{mnih2015human} because of scalability to environments with a large number of states.
In our grid world environment (Fig.~\ref{fig:grid}), there are more than 8 billion states.\footnote{Each state is a tuple of 16 elements consist of robot and each of the items' (door key, green and purple cube) positions, state of the lamp and fire (on or off), and state of the door (open or close).}
The algorithm takes the STL specification of the task $\varphi_{task}$, the goal state and the hyper-parameters (i.e., $M$, $C$, $\gamma$, and etc.) as input, and generates the optimal policy that respects $\varphi_{task}$ as output. 
The main RL loop runs for $|episodes|$. 
In each episode, first, the state is set to initial state and the partial trajectory is set to $\emptyset$. 
While the robot has not reached the final state or maximum number of states is not reached, the robot explores the grid environment and the reward is computed as robustness of the partial trajectory with respect to $\varphi_{task}$. 
The robot experiences are recorded in replay memory to be used later for training the $Q$ network. 
Whenever the replay buffer size exceeds $M$, we start training the $Q$ network using the bellman equation. 
We update the weights of target action-value function $\hat{Q}$ with the weights of $Q$ in every $C$ episodes. 
For our running example, with $\varphi_{task} = \F_{[0,15]}(\atom{lampOn} \aand \F_{[0,10]}(\atom{itemOnRobot(purpleCube)}))$, the reward converges in less than 15000 episodes, and the learned policy is illustrated in Fig.~\ref{fig:policy}.

\section{Experiments}
\label{sec:experiments}
To evaluate \framework{}, we qualitatively and quantitatively examine its performance in our gridworld (Fig.~\ref{fig:grid}) on natural language specifications whose underlying STL formulas exhibit various qualities. 

Table \ref{tab:results} shows a set of manually curated natural language sentences describing a variety of tasks and user constraints paired with the most frequent STL formulas ultimately predicted by \framework{}.
We use the GPT-3 based paraphrase generator to generate 107 paraphrases each for the sample sentences in Table \ref{tab:results}.
We also design an Oracle user to interact with \framework{} to answer posed interaction questions.
The Oracle user is a simple, rule-based program that provides the correct answer to any given question about the true, underlying STL formula.
Table~\ref{tab:results} shows the source language input, before paraphrasing, the number of provided demonstrations, and the \framework{} average number of enumerated formulas, user interactions, success rate, and run-time.\footnote{The time for loading the model, initializing the tools and getting the demo from the user are excluded from run-time. The run-time only measures the main part, i.e., converting natural language to STL. We run the experiments on an Intel Core-i7 Macbook Pro with 2.7 GHz processors and 16 GB RAM.}

\newcommand{\tablenlsize}[1]{{\tiny #1}}
\newcommand{\tabletypesize}[1]{{\small #1}}
\newcommand{\tabletyperot}{0}

\definecolor{Gray}{gray}{0.90}
\newcolumntype{a}{>{\columncolor{Gray}}r}
\newcolumntype{g}{>{\columncolor{Gray}}p{4cm}}

\begin{table*}[t]
  \centering
  \setlength{\aboverulesep}{0pt}
  \setlength{\belowrulesep}{0pt}
  \begin{tabular}{lgarrrrrp{6cm}}
  & \multicolumn{2}{c}{\bf ---User Input---} & \multicolumn{5}{c}{\bf ---\framework{}---} \\
  Type & Pre-paraphrase Natural Language & \#Ds & \#EFs & \#UIs & SR & RT & Most Frequent STL Prediction \\
  \toprule
  \addlinespace[-\aboverulesep]
  \multirow{2}{*}{\rotatebox[origin=c]{\tabletyperot}{\tabletypesize{C}}} &
    \tablenlsize{Always don't hit into walls.} 
        & 2 
        & 4.2 
        & 1.0
        & \phantom{0}80\%
        & 2.75 
        & \tablenlsize{$\alw_{[0,1000]}(\neg \atom{robotAtWall})$} \\
    
    & \tablenlsize{Always do not walk into water.}
        & 2
        & 4.9
        & 1.0
        & \phantom{0}90\%
        & 2.73 
        & \tablenlsize{$\alw_{[0,1000]}(\neg \atom{robotAtWater})$} \\
    
    \cline{1-8}
    \multirow{2}{*}{\rotatebox[origin=c]{\tabletyperot}{\tabletypesize{S}}} &
    \tablenlsize{Pick up the purple cube.}
        & 1
        & 2.0
        & 1.1
        & 100\%
        & 2.46 
        & \tablenlsize{$\F_{[0,15]}(\atom{itemOnRobot(purpleCube)})$} \\
     
    & \tablenlsize{Turn off the fire.}
        & 1
        & 2.0
        & 1.1
        & \phantom{0}80\%
        & 2.31 
        & \tablenlsize{$\F_{[0,8]}(\atom{fireOff})$} \\
     
    \cline{1-8}
    \multirow{4}{*}{\rotatebox[origin=c]{\tabletyperot}{\tabletypesize{Q}}} &
    \tablenlsize{Open the door and then charge yourself.}
        & 1
        & 3.1 
        & 3.0
        & \phantom{0}90\%
        & 3.28 
       & \tablenlsize{$\F_{[0,10]}(\atom{doorOpen} \aand \F_{[0,15]}(\atom{chargerPlugged}))$} \\
        
    & \tablenlsize{Go to location (7, 4) and pick up the green cube.}
        & 1
        & 2.7 
        & 3.1 
        & \phantom{0}72\%
        & 2.87 
       & \tablenlsize{$\F_{[0,15]}(\atom{robotAt(7,4)} \aand \F_{[0,4]}(\atom{itemOnRobot(greenCube)})$} \\
        
    & \tablenlsize{Turn on the lamp before picking up the purple cube.}
        & 1
        & 45.1 
        & 3.1 
        & \phantom{0}77\%
        & 5.79
        & \tablenlsize{$\F_{[0,10]}(\atom{lampOn}\; \Until_{[0,8]}\; \atom{itemOnRobot(purpleCube)})$} \\
         
    & \tablenlsize{Open the gate before picking up the green cube.}
        & 1
        & 26.9 
        & 3.0
        & \phantom{0}71\%
        & 14.76
        & \tablenlsize{$\F_{[0,6]}(\atom{doorOpened}\; \Until_{[0,6]}\; \atom{itemOnRobot(greenCube)})$} \\
        
    \cline{1-8}
    \multirow{2}{*}{\rotatebox[origin=c]{\tabletyperot}{\tabletypesize{M}}} &
    \tablenlsize{Turn on the lamp or turn on the fire.}
        & 2
        & 2.0
        & 1.1 
        & \phantom{0}81\%
        & 4.48
        & \tablenlsize{$\F_{[0,20]}(\atom{lampOn} \vee \atom{fireOn})$} \\
        
    & \tablenlsize{Sit on the chair or pick up the purple cube.}
        & 2
        & 2.0
        & 1.1
        & 100\%
        & 4.29
        & \tablenlsize{$\F_{[0,20]}(\atom{robotSittingOnChair} \vee \atom{itemOnRobot(purpleCube)})$} \\
        
        
        
    \bottomrule
  \end{tabular}
  \caption{{\rm \framework{} performance on sample natural language inputs across 107 GPT-3 paraphrases of the inputs for fixed user demonstrations per row (\#Ds).
  We report the number of enumerated formulas (\#EFs), average user interactions (\#UIs) to select a final formula, success rate (SR) of finding the exact match correct formula, and average runtime in seconds(RT). 
  The task types include (C)onstraint, (s)ingle, se(Q)uence and (M)ultiple-choice.}}
  \label{tab:results}
  \vspace{-20pt}
\end{table*}

\subsection{Results Across Description Types}
\mypara{User Constraints}
A human (Oracle) provides the general constraints for a task, and one positive demonstration that satisfies the constraint and negative one that does not satisfy or violates it.
The human user is asked ``For how many seconds do you want the constraint to be satisfied?", and answers ``1000 seconds'', and so the best STL formula is predicted as $\varphi = \alw_{[0,1000]}(\neg \atom{robotAtWall})$: ``Always, in the next 1000 seconds, the robot should not run into walls".

\mypara{Single Tasks} A human user provides a single task, such as ``Pick up the purple cube", and a positive demonstration of the task.
Negative examples are generated from this positive example based on the principle of ``no excessive effort".
The user is asked about timing requirements, such as: ``In how many seconds, should the robot complete the task?", and in this case answers ``15 seconds''.
The formula $\varphi = \F_{[0,15]}(\atom{itemOnRobot(PurpleCube)})$ is predicted, which means ``In the next 15 seconds, the purple cube should be picked up by robot".


\mypara{Sequence of Tasks}
A sequential task, such as ``Go to location (7, 4) and pick up the green cube", requires the robot to do one thing before another---a temporal dependency.
The STL formulas that do not give such guarantee can be eliminated from the candidate formulas, and in this case \framework{} predicts $\F_{[0,15]}(\atom{robotAt(7,4)} \aand \F_{[0,4]}(\atom{itemOnRobot(greenCube)})$: ``In the next 15 seconds, the robot should reach to location (7, 4) and after robot reaches (7, 4), it should pick up the green cube in the next 4 seconds".

Another example of a sequential task is ``Turn on the lamp before picking up the purple cube". 
The word ``before" implies the temporal dependency, and the formula predicted is $\F_{[0,10]}(\atom{lampOn} \Until_{[0,8]} \atom{itemOnRobot(purpleCube)})$: ``Turn on the lamp" happens in the past of ``pick up the purple cube".

\mypara{Multiple-choice Tasks} ``Turn on the lamp or turn on the fire" means that the robot is required to complete at least one of the two tasks. 
In such cases, the user can provide two positive demonstrations showcasing the alternative goals. 
The STL formula predicted for this example is $\F_{[0,20]}(\atom{lampOn} \oor \atom{fireOn})$: ``In the next 20 seconds, the lamp should be on or the fire should be on".

\section{Related works \& Conclusions}
\label{sec:related_works}
Prior work has used STL for reinforcement learning applications. 
Quantitative semantics of STL can be used as reward functions for evaluating robotic behaviors~\cite{balakrishnanstructured}.
STL formulas can be used to rank the quality of demonstrations in robotic domain and also computing the reward for RL problems~\cite{puranic2021learning}.
However, those works put the burden of specifying the correct STL formulas on users, and can require $3x$ more demose than \framework{} despite using a similar environment~\cite{puranic2021learning}.

There has been a tremendous effort in learning temporal logic languages from natural human languages~\cite{gavran2020interactive, he2021english, nelken1996automatic, ranta2011translating, kress2008translating, autili2015aligning}.
These works variously assume a particular format for natural language, are limited to a specific fragment of formal logic, or have scalability and robustness issues. 
Our work addresses these shortcomings by operating over the space of all possible STL formulas, leveraging interaction with the user to repair ambiguities, and scaling to a larger state space.


\bibliographystyle{unsrtnat}
\bibliography{references}

\end{document}